# AI-enhanced conversational agents for personalized asthma support
## Factors for engagement, value and efficacy

Laura, L.M., MORADBAKHTI*

Dyson School of Design Engineering, Imperial College London, United Kingdom, l.moradbakhti@imperial.ac.uk

DORIAN, D.P., PETERS

Dyson School of Design Engineering, Imperial College London, United Kingdom, d.peters@imperial.ac.uk

JENNIFER K., J.K.Q., QUINT

Faculty of Medicine, National Heart & Lung Institute, Imperial College London, United Kingdom, j.quint@imperial.ac.uk

BJÖRN, B.S., SCHULLER

Department of Computing, Imperial College London, United Kingdom, bjoern.schuller@imperial.ac.uk

DARREN, D.C., COOK

Dyson School of Design Engineering, Imperial College London, United Kingdom, darren.cook@imperial.ac.uk

RAFAEL A., R.A.C., CALVO

Dyson School of Design Engineering, Imperial College London, United Kingdom, r.calvo@imperial.ac.uk

Asthma-related deaths in the UK are the highest in Europe, and only 30% of patients access basic care. There is a need for alternative approaches to reaching people with asthma in order to provide health education, self-management support and bridges to care. Automated conversational agents (specifically, mobile chatbots) present opportunities for providing alternative and individually tailored access to health education, self-management support and risk self-assessment. But would patients engage with a chatbot, and what factors influence engagement? We present results from a patient survey (N=1257) devised by a team of asthma clinicians, patients, and technology developers, conducted to identify optimal factors for efficacy, value and engagement for a chatbot. Results indicate that most adults with asthma (53%) are interested in using a chatbot and the patients most likely to do so are those who believe their asthma is more serious and who are less confident about self-management. Results also indicate enthusiasm for 24/7 access, personalisation, and for WhatsApp as the preferred access method (compared to app, voice assistant, SMS or website). Obstacles to uptake include security/privacy concerns and skepticism of technological capabilities. We present detailed findings and consolidate these into 7 recommendations for developers for optimising efficacy of chatbot-based health support.

## 1 INTRODUCTION

Eight million people (12% of the population) in the UK are estimated to have a diagnosis of Asthma [22]. Despite effective treatments and extensive efforts, the UK death rate from asthma is the highest in Europe. Moreover, 65% of people with asthma in the UK do not receive the professional care they are entitled to (e.g., a yearly review, feedback on how to use their inhaler correctly, or an asthma action plan) [35]. In the latest Asthma+Lung UK survey [3], it was reported that only 30% of Asthma patients received basic care in 2021, the lowest proportion since 2015. Moreover, according to the same report, almost half (44%) of the patients who are admitted to hospital did not receive follow-up care, with younger adults (18-29) receiving the lowest level of basic asthma care [2].

One of the reasons for this is believed to be poor self-assessment of risk by those with asthma. For instance, according to Asthma+Lung UK, the illness is often not taken seriously enough as reports show that 1 in 6 people in the UK do not know, or are not sure, if asthma can be fatal [1]. Moreover, health education and self-management behaviours play a critical role. In the UK, Asthma associated care costs are at least £1.1bn per year [21] and the bulk of the burden centres on differences in asthma control rather than severity. People with severe uncontrolled Asthma (SUA) are prone to more symptoms, night-time awakenings, rescue medication use, and worse self-reported outcomes [9] and are estimated to represent 3-times the cost to the healthcare system as compared with patients with severe but controlled disease [9]. Studies have shown, informed self-management that is accompanied by professional reviews improves asthma control, reduces exacerbations, and improves overall quality of life [25].

---

* Corresponding author

These statistics demonstrate an urgent need for alternative approaches to reaching people with asthma in order to: a) improve risk self-assessment, b) improve health literacy and self-management and c) encourage them to seek traditional care as needed. One way to support people with asthma could be through a specialised conversational agent. A Conversational Agent (CA) can be any type of dialogue system that has natural language processing capabilities and can respond to input using human language. Moreover, the input/output is not restricted to text but could also happen via voice. Although CAs can be embodied avatars or physical robots, for the purposes of the current study, we refer only to CAs that are non-embodied chatbots used via applications such as WhatsApp.

New generative chatbot technologies, like ChatGTP, are rapidly becoming pervasive and people are likely to continue to consult them for healthcare advice despite safety risks [11,20,33]. Conversational interfaces based on Large Language Models can be intuitive, engaging, and provide a level of personalisation in their responses based on user input. However, these technologies rely on probabilistic models, trained on information from across the internet, in order to generate answers that have never been vetted. These systems are known to fabricate information and present it confidently (sometimes adding artificial citations to sources). Such 'hallucinations' could be dangerous or even life-threatening when the system is responding to someone's medical questions. As such, misinformation remains a significant problem for publicly available generative AI services rendering them unsafe for health applications. The development of a trustworthy conversational system must involve health experts (including patients and doctors) and must exclusively provide health information that has been verified as safe and accurate for the specific context.

There is existing evidence that mobile technology can be used effectively for health interventions to help people learn about their illness, complete self-assessments, track symptoms and be directed to appropriate care options. Several bespoke apps for asthma [14,16,26,34], and a smaller number of studies on conversational agents designed for children and adolescents have been reported in the literature [15,31]. However, research is missing that might guide an effective mobile conversational agent for adults with asthma. While the technology itself holds promise, successful outcomes are entirely contingent on human acceptance and use, and research is lacking on what adult patients' concerns, expectations, preferences, and motivations would make such a conversation-based health intervention valuable to them. Furthermore, how these needs might vary for the groups that most stand to benefit from alternate interventions (e.g., patients with poor asthma control, with lower health literacy, or who do not access traditional health services) is not clear.

There are many open questions to do with the potential for meaningful patient engagement with a conversation-based asthma intervention. For example, what would patients be looking for from conversational health support? What would be necessary for them to value and trust a technology enough to use it? How does existing technology use play a part? While CAs have a lot of potential to help people with asthma, they are a relatively new type of technology in the public sphere and, in addition to common concerns about privacy, people may mistrust a CA's ability to help. Also, there is not enough understanding of what motivates or hinders patients' use of CAs in general, or how socio-economic variables may relate to lack of trust in, or engagement with, CAs.

The study described herein aims to provide answers to these questions and contribute a better understanding of the features required to make a conversational agent for asthma successful based on the expectations, needs, and motivations of adult asthma patients. Additionally, attention is given to different groups of patients and how developers might most effectively meet the needs of those who stand to benefit the most from such digital services.

## 2 RELATED WORK

This section describes previous work in the three intersecting areas relevant to this project: mobile technologies for health broadly and asthma specifically; *conversational* mobile technologies for health broadly and asthma specifically; and the use of WhatsApp as a platform for conversational mobile digital health.

### 2.1 Mobile Technologies for Asthma

Mobile technologies focusing on healthcare support have been proven to be cost-effective across health domains [35]. Specifically, there is evidence that mobile technology can be used effectively to help patients learn about their illness, complete self-assessments, track symptoms and be directed to appropriate care options. According to a systematic review, 87% improved adherence and 53% improved health outcomes for users [27]. Moreover, they are a cost-effective way to reach patients in a personalised and time-efficient manner, for example, by reminding patients to engage in self-management tasks (e.g., medication adherence, electronically accessible actions plans etc.) [35]. With respect to asthma specifically, bespoke apps are the most common approach to mobile interventions for asthma. For example, a number of bespoke apps for young people [16,27] have demonstrated positive outcomes for improving self-management. For adults the ASTHMAXcel mobile application was designed to improve asthma knowledge and outcomes [13]. The app was tested in multiple studies and revealed an increase in asthma knowledge [8,13] as well as a decrease in clinical outcomes such as emergency department visits, hospitalizations, and Prednisone utilization [12] amongst patients.

**2.2 Conversational Agents for Health**

Previous studies have shown that Cas can help patients with a range of healthcare tasks, such as delivering clinical and health risk information, providing elements of basic care, and supporting behaviour change [30] through personalized dialogue. The use of conversational agents for health has been extensively reviewed [19,23,32].

With respect to asthma, specifically, three studies explored the use of CAs for young people. Rhee et. al. [28] described an automated mobile phone-based asthma self-management aid for adolescents, able to interpret conversational English SMS messages describing asthma symptoms. The system was tested over two weeks by 15 adolescent-parent dyads, and response rates for daily messages were 81%–97% for the adolescents. Results showed that, following the intervention, participants' awareness for both symptoms and triggers, their sense of control, and treatment adherence were higher, with benefits to the partnership between parents and adolescents.

Kadariya et. al. [15] tested kBot, a chatbot capable of interacting both via text and voice in the form of an Android application, designed to support paediatric asthma patients (age 8-15). Overall, kBot received good technology acceptance and system usability scores from clinicians (N = 8) and researchers (N = 8) but no patient testing was reported.

More recently, a study by Kowatsch et al. [17] tested MAX, a conversational agent that was designed to increase knowledge of asthma and behavioural skills (such as inhalation technique) amongst 10–15-year-olds with asthma. MAX incorporated different modes of communicating with the CA, for example, care professionals could communicate with it via email, patients via a mobile app and family members via SMS text messages. The study results showed, not only high acceptability of the CA by all stakeholders, but also improved cognitive and behavioural skills. These studies with children and adolescents suggest promise for conversation-based digital health care. However, they are limited to tests with small groups of paediatric patients.

Messaging platforms are widely used across age groups in the UK and elsewhere: According to Statista [18] 80% of 16-64 old in the UK use WhatsApp and 59.5% used Facebook Messenger. As such, the potential for conversation-based support for adults could be equally promising, but we need a better understanding of adult users and what might drive them to engage with conversational tools in the real world, including how this may vary between different groups with respect to asthma severity, control, and other factors.

**2.3 WhatsApp and Digital Health**

Although conversational agents are often delivered as part of custom apps, they can also leverage existing and familiar chat-based tools such as SMS, WhatsApp, or web-based messaging systems. Indeed, we began our project with the intention of creating a bespoke app, but early engagement with patients and patient advocates led us to change this plan and shift to the widespread platform, WhatsApp. As mentioned, most of the UK adult population already use WhatsApp which is probably why it was favoured by our patient participants. Using a familiar and widespread platform also meant we could remove the additional friction and technical barriers associated with bandwidth, storage capacity, usability, learnability, and concerns over trustworthiness that can come with downloading and using a new app for the first time.

Previous research has demonstrated positive results for the use of WhatsApp in healthcare settings. For example, an online questionnaire assessing patient satisfaction with WhatsApp for sharing health information (collected from 14 WhatsApp asthma and allergy groups), concluded that WhatsApp is an effective social media program for improving patient knowledge and behaviour [29]. Furthermore, a study conducted by Asthma UK offered young adults access to an asthma nurse who would assist them in managing their asthma via WhatsApp [10]. The service was well received with 67.1% of participants reporting improved confidence in their asthma management post usage.

Similarly, a study conducted in Latin America confirmed an interest in WhatsApp as a tool to receive information about asthma irrespective of age (mean age 43), as 61.5% of patients in the study reported the highest interest in WhatsApp (as compared to SMS, Facebook, Twitter, and E-mail) as a tool for communication [5]. It is noteworthy that patients reported SMS as the most used service (69.9%) while still indicating that WhatsApp would be the most preferred tool for Asthma communication. WhatsApp has several advantages over SMS which may help explain its penetration. It is often more cost effective than SMS because SMS messages are charged differently by mobile network providers than web-based messaging platforms like WhatsApp are; unlike with SMS, the geographic reach of WhatsApp is global (i.e. not limited to the mobile service region); and it provides better support for group communication, audio, video and the option to retain an account even when one's phone number changes.

Despite the above evidence for the potential benefits of using WhatsApp, it is not as widely used as websites or custom apps. One reason (which applies to all messaging platforms) is that text and voice conversations do not provide as many affordances (i.e. visual displays and menus interface) as other media. Another reason, more specific to WhatsApp, involves concerns over privacy. For example, in order to receive research ethics approval for a follow-up study, the research team carried out thorough discussions with both the ethics committee and the university Data Protection Office over the use of WhatsApp. Some of the concerns raised and how we recommend addressing them are listed below:

- **Sensitive data**. A health app could include sensitive data. Given these concerns, we recommend not asking users for any sensitive information, and only asking for the minimum data required to provide functionality.
- **Data use by Meta**. Because WhatsApp is owned by Meta (previously Facebook), there is no way of knowing how they will use data from chatbot conversations. To address these concerns, we recommend including an explanatory section specific to WhatsApp in the participant information sheet. It's also important to note that, as of September 2023, WhatsApp has end-to-end encryption, meaning the conversation is only readable by the two 'ends' communicating, in this case the patient and our servers. This means, not even WhatsApp can read them. As such, the onus is on the entity hosting the chatbot servers to manage data securely and ethically rather than on Meta.
- **Local storage vulnerabilities**. Local backups of the conversations made on a person's phone could be compromised if the phone is lost. To address these concerns, we recommend instructing users on how to turn off the backup if they wish.
- **Management of identifiable data**. To further address privacy, we propose it is not necessary to ask for any identifiable information, and to use only the WhatsApp phone number to identify a user within the system. Conversationally, the user can be invited to give a nickname rather than their real name. For a data analysis team, data can be anonymised further (phone numbers and nick names can be removed).

## 2.4 Summary

The studies described above, including the asthma chatbot studies with children and adolescents, the survey studies exploring the potential for use of WhatsApp in health, and the significant evidence for the efficacy of mobile interventions more broadly within healthcare, all provide promising evidence for a new conversation-based digital health care approach to complement primary and secondary care. These are summarised in Table 1 to clarify visually how existing work is distributed with respect to which technology it refers and whether it refers to health broadly or asthma specifically.

Table 1: Overview of Categories Under Which the Related Work Studies Fall

|  | for Health (broadly) | for Asthma (specifically) |
|---|---|---|
| **Mobile Technology (broadly)** (e.g., apps, bespoke etc.) | Ramsey et al., 2019b [27] <br><br> Whittamore, 2017 [35] | MAX (Kowatsch et al., 2021) [17] <br><br> ASTHMAXcel (Hsia et al., 2020) [12,13] |
| **WhatsApp** (specific chat technology) | Saeed Tayeb, 2019 [29] | Calderón et al., 2017 [5] <br><br> Cumella et al., 2020 [10] |
| **Conversational Agents** - Automated Chatbots that can be used through chat technologies) | Singh et al., 2023 [30] | kBot (Kadariya et. al., 2019) [15] <br> Rhee et al., 2014 [28] |

What remains to be explored is a nuanced understanding of what adult users themselves would use and value, what might drive them to engage with conversational health tools in the real world, and how this might differ for diverse groups (e.g., based on differences in disease severity, symptom control, confidence with technology or trust in the healthcare system). For this study, we specifically sought this improved understanding for the population of adults with asthma within the UK, and we did so with a view to improving asthma risk self-assessment, asthma health literacy and access to care. As such, our study aimed to answer the following research questions:

1. To what extent do different adults with asthma in the UK trust the healthcare system and does this level of trust correlate with interest in using a chatbot for asthma?
2. Are those people who are more interested in using an asthma care chatbot more or less likely to already be accessing basic health care services (i.e., GP)?
3. In what ways would adults with asthma prefer to access an asthma care chatbot?
4. What features, characteristics and style preferences for a chatbot would maximise the likelihood of adults with asthma engaging meaningfully with it?
5. To what extent does a person's confidence using technology affect their willingness to use a chatbot for asthma?
6. Is there a relationship between an individual's interest in using an asthma care chatbot and their self-assessment of the seriousness of their Asthma?

## 3 METHODS

### 3.1 Overview

The survey study described herein is part of a larger overarching project funded by the UK Engineering and Physical Sciences Research Council (EPSRC) and in partnership with Asthma+Lung UK, an asthma charity organisation. A description of the overarching project is available in the protocol paper by Calvo et al. [7].

### 3.2 Recruitment

Participants were recruited through YouGov [38]. To fulfil our recruitment criteria, we only recruited participants with self-reported asthma from the YouGov database. Participants were paid based on YouGov rates.

### 3.3 Survey

A survey with 38 closed and 12 open-ended questions was conducted with adults living with asthma to gain a better understanding of how design and demographic features might influence interest in, and preferences for, a chatbot to support asthma management. A full list of the survey questions is attached in the supplementary materials. The survey questions were approved by the Imperial College London ethics committee (#21|C7403).

### 3.4 Analysis

The quantitative results were analysed using SPSS v28 [36] to run all statistical analyses of the quantitative survey results. Firstly, descriptive statistics were analysed to gain a better understanding of the sample and participants' general preferences for an asthma chatbot. Following the descriptive statistics, further analyses were conducted with a focus on users' trust in the healthcare system and to highlight differences between participants who were and were not interested in the chatbot. For all analyses the alpha level was set at .05.

Qualitative analysis of the open-ended free text responses in the survey was conducted via a process of inductive thematic analysis following the steps outlined in Braun & Clarke [4] and using NVIVO (v11 for Mac) [37]. The aim was two-fold: 1) to identify more detailed insights underlying quantitative answers to the closed-ended questions and 2) to highlight patterns of responses or themes salient across participants' free-text contributions. In other words, text responses were coded iteratively, and themes generated were based on repeated patterns of ideas shared across participants. Resulting themes highlight some of the more common motivations and rationale for the quantitative results.

## 4 RESULTS

### 4.1 Quantitative Results Overview

A total of 1257 people with self-reported asthma completed the survey (43% male, 57% female, 4 participants left the gender question blank). 47.8% of participants reported that their asthma was diagnosed under the age of 16, while 51.7% of participants reported that their asthma was diagnosed over the age of 16 and 0.5 % of participants preferred not to provide this information. 98.9 % of participants received their diagnosis more than 6 months ago, 0.8% of participants preferred not to say when their diagnosis happened, and 0.3% of people were diagnosed with asthma within the last 6 months at the time of participation in the study.

Most participants were over 40, specifically 44% were older than 55, 30% between 41-55, 19% between 31-40 and 7% between 18-30. Most participants had either completed a university degree (48%) or trade/technical/vocational training (19%) while 33% indicated that secondary school was their highest education level. For 0.4% the highest education level was the completion of primary school. 56% of participants spent most of their life in a medium to small town, 23% in a rural area and 21% in a big city.

67 participants (5.3 %) from our sample identified as being part of a minority ethnic group, while 1178 participants (93.7%) did not and 12 participants (1%) preferred not to disclose. The number of participants who identified as part of a minority ethnic group was too small to allow for *comparisons* between participants who do and do not identify as such, however, some *correlations* with minority group identification and other variables are reported.

### 4.2 Descriptive Results

YouGov shares some data they collect from all their registered users. Existing YouGov data on WhatsApp use was available for 1011 of our participants and showed that 85% use WhatsApp. According to our own questions, 74% of participants use smartphone messaging apps daily and 60% specifically use WhatsApp *on a daily basis*.

In contrast, 92.5% of participants indicated that they had never used a mobile app to help with their Asthma (e.g., to track their symptoms). Instead, participants were more likely to turn to existing familiar technologies (i.e., internet search) or to trusted people (nurses, doctors or family) for asthma information. The graph below shows which platform/source of information participants currently utilise to get information about Asthma (multiple answer options, total $N = 1257$).

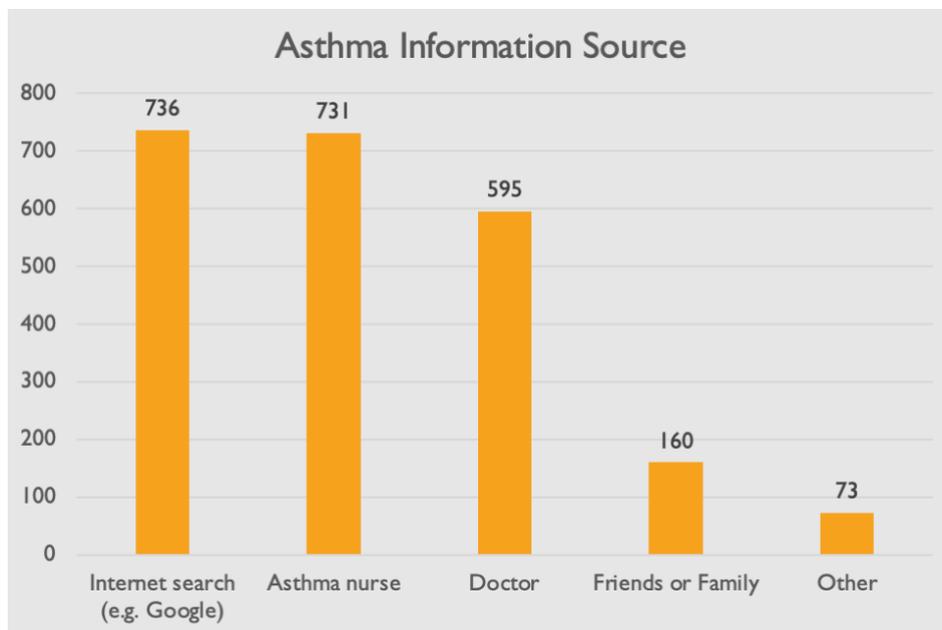

Figure 1: Overview of Channels Participants Currently Use to Receive Information about Asthma.

Participants were asked about which platform participants would prefer to use to chat with an asthma chatbot and they were allowed to select more than one. The majority indicated a preference for WhatsApp (see Figure 2 below), closely followed by a website.

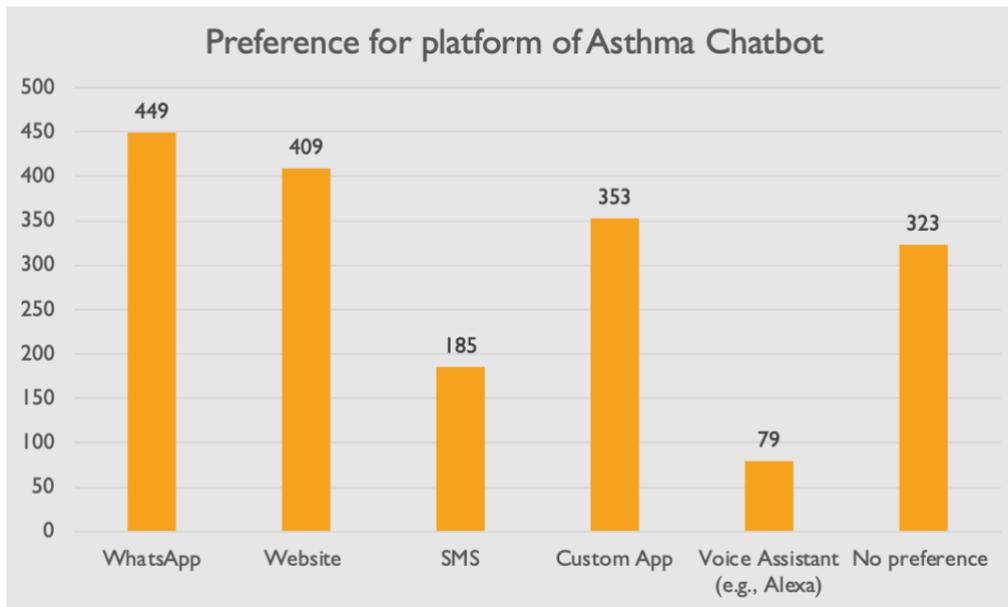

Figure 2: Overview of Participants' Preferred Platform to Use the Asthma Chatbot.

Overall, the majority of participants (53%) indicated that they would indeed be interested in using an asthma chatbot via a messaging service like WhatsApp (17% very interested, 36% somewhat interested).

An even higher number of participants 59% (22% very interested, 37% somewhat interested) indicated interest in an asthma chatbot that could detect asthma severity from sound of voice-- a technological feature currently in development by our research team. For this question, we asked participants: *Imagine AsthmaBot could detect your asthma severity by listening to the \*sound of your voice\*. You would speak into your phone to provide a short voice recording and AsthmaBot would analyse this recording for signals (that only a computer can detect) to determine your asthma severity. How interested would you be in trying this voice feature?*

**4.3 Trust in Healthcare System**

To analyse participants' trust towards the healthcare system, Spearman rank-order correlations were conducted. There was a significant negative correlation between participant education level and their trust in the healthcare system, $r_s(1255) = -.07$, $p = .014$. That is, the higher a participant's educational attainment, the more likely they were to trust the healthcare system.

Participants' confidence in technology significantly positively correlated with their trust in the healthcare system: $r_s(1255) = .190$, $p < .001$. In other words, participants who were more confident in their ability to use technology also had higher trust in the healthcare system.

Participant trust in the healthcare system just fell short of a significant negative correlation with their identification as a minority ethnic group (or not): $r_s(1255) = -.054$, $p = .056$. Therefore, we can only report that there is a tendency towards participants who do not identify as part of a minority ethnic group to trust the healthcare system more than people who do identify as an ethnic minority. This would be an important question to explore in a future study with a larger sample and would support previous research on experiences of discrimination and distrust in the healthcare system[24].

Age correlated significantly and negatively with trust in the healthcare system $r_s(1255) = -.07$, $p = .009$ indicating that older participants were more likely to trust the healthcare system than younger ones.

Participants' confidence in dealing with their asthma significantly correlated with participants' trust in the healthcare system $r_s(1255) = .101$, $p < .001$, showing that participants who were confident in managing their asthma trusted the healthcare system more. This may suggest that positive experiences with the healthcare system led to better management and therefore, confidence in ability to manage their asthma. However, there are many possible explanations for such as result: for example, people who trust the system less may not access healthcare and therefore may not get what they need to manage their asthma well; people who have had negative experiences with the system may not have gotten what they need to manage their asthma well despite accessing it; or people with asthma that is more difficult to treat and manage may more clearly see the limitations of the healthcare system.

### 4.4 Comparing participants interested v. not interested in an asthma chatbot.

To allow a straightforward interpretation of the data, some of the variable answers were grouped. The example below (for the variable: 'How interested would you be in using Asthma Bot?') shows how the Likert-scale answers were grouped into a binary format (see Table 2).

Table 2: Categorical Variable Split between Participants Interested in the Asthma Chatbot and Not Interested in the Asthma Chatbot.

| Interested | | Not interested | | |
|---|---|---|---|---|
| Very Interested | Somewhat Interested | Neutral | Not particularly interested | Not at all interested |

There was no significant correlation between participants' interest in using an asthma chatbot and either their age, education level, or identification as a minority. This suggests that neither age, education level nor identification with a minority ethnic group is likely to be a barrier to engagement.

### 4.5 Healthcare Utilisation Predicts Interest

We conducted a CHI Square test for association to measure the relationship between participants' trust in the healthcare system and their interest in an asthma chatbot. The relationship is significant $\chi2(1) = 9.75$, $p = .002$, $\varphi = .088$. Overall, the majority of participants trusted the healthcare system (83%) while 17% did not. Out of the participants who were not interested in an asthma chatbot, 20.1% indicated they did not trust the healthcare system, while 79.9% did. Out of the participants who were interested in using an asthma chatbot, a higher percentage trusted the healthcare system (86.5%) and a lower percentage of participants did not trust the healthcare system (13.5%). This indicates that people who are interested in an asthma chatbot were more likely to trust the healthcare system than those who are not interested.

To check for a difference between participant interest in an asthma chatbot and their current support from a GP, we conducted another CHI Square test for association. The relationship is significant $\chi2(1) = 12.72$, $p < .001$, $\varphi = .101$. Overall, 20.5% of participants do not seek support from a GP, while 79.5% do. Of the participants who were interested in the chatbot, 16.7% did not already get support from a GP, while 83.3% do. For participants who were not interested in the chatbot, 24.8% did not get support from a GP, while 75.2% do. This suggests that overall participants with an interest in the asthma chatbot were also more likely to be seeking support from a GP. This aligns with our results that show that people who see their asthma as more serious are more likely to be interested in a chatbot. Moreover, participants' perception of the seriousness of their asthma positively and significantly correlated with the ACQ questions: difficulty sleeping due to asthma $rs(1255) = .313$, $p <.001$; asthma symptoms during the day $rs(1255) = .279$, $p <.001$; asthma affecting usual activities $rs(1255) = .361$, $p <.001$; asthma caused A&E visit in the last two years $rs(1255) = .275$, $p <.001$. Taken together, these findings suggest that, at least some of the participants not interested in a chatbot lack interest because they don't feel their asthma is serious enough to motivate use. Qualitative results provide additional evidence for this conclusion (see Section 4.10).

### 4.6 Communication Preferences: Participants Would Value Asthma Voice Recognition But Still Favour Human Support

Regarding participants' interest in a feature that detects asthma severity through the sound of one's voice, voice recording could be more appealing (e.g., novelty, perceived accuracy) or less appealing (e.g., privacy concerns) than the standard text-based chatbot. Therefore, we analysed the relationship between interest in such a feature and general interest in an asthma chatbot with a CHI Square test for association. Again, both variables were coded as interested vs. not interested (see Table 2). The relationship is significant $\chi2(1) = 347.100$, $p < .001$, $\varphi = .525$. Overall, 59.7% of participants were interested in the voice feature, while 40.3% were not. Out of the participants who were interested, 84.1% were interested in the voice feature, while 15.9% were not. Out of participants who were not interested in an asthma chatbot, only 32.4% were interested in the voice feature and 67.6% were not. These differences show that most participants with a general interest in an asthma chatbot were also interested in a voice feature to detect severity, and that a subgroup of participants not interested in an asthma chatbot, were however, interested in voice-based severity detection for asthma.

A CHI Square test for association was conducted to test the relationship between participants' preference for talking to a human about their asthma and their interest in an asthma chatbot. The relationship is significant $\chi2(1) = 17.05$, $p < .001$, $\varphi = .116$. Overall, 85.5% of participants reported that they would prefer to talk to a human about their asthma. Out of the participants interested in an asthma chatbot, 81.7% would prefer to talk to a human, while 18.3% didn't. Out of the participants who were not interested in an asthma chatbot, an even higher percentage had a preference to talk to a human: 89.9%, while only 10.1% did not. The findings demonstrate that participants who were not interested in using a chatbot for asthma, unsurprisingly, have a stronger preference for in-person conversations about their asthma.

It is also worth noting that while the large majority would prefer to speak to a human about their asthma, a similar majority had sought asthma information via the internet. This suggests that, although human interaction is preferred, where it is not available or easily accessible, most people turn to non-human options for support.

### 4.7 Conversational Style: Most prefer a chatbot that is friendly, reassuring and like a nurse.

Research suggests that defining personality dimensions and a clear conversational style to guide the dialogue of a chatbot, provides a more consistent and effective user experience [6]. Chatbot conversational style can vary across different personality dimensions, for example, it might be more or less formal/informal, authoritative/submissive or convey warmth via word choice. In order to determine the kind of conversational style most people with asthma would feel comfortable interacting with in a chatbot, the survey provided a list of personality descriptors and asked participants to select those they believe to be most appropriate for an asthma chatbot. Most respondents preferred an asthma chatbot that was 'reassuring', friendly' and 'like talking to a nurse'. It's notable however, that a significant number selected other preferences, for example, favouring a 'direct' approach or an experience that was more 'like talking to a doctor', suggesting different styles suit different people. These results are summarised in Table 3.

Table 3: Patients who are interested in Asthma Bot and their preferred conversational style when communicating with a virtual assistant, the top 8 most popular choices.

| Reassuring | Like talking to a nurse | Friendly | Caring | Good listener | Like talking to a doctor | Direct | Informal |
|---|---|---|---|---|---|---|---|
| 343 | 335 | 316 | 261 | 198 | 196 | 164 | 147 |

### 4.8 Technology Experience Predicts Interest

To assess the relationship between participants' confidence in using technology and their interest in an asthma chatbot, we conducted a CHI Square test for association. The relationship is significant $\chi2(1) = 13.73$, $p < .001$, $\varphi = .105$. Overall, the large majority of participants were confident in their use of technology (95%). Out of the participants who were not interested in an asthma chatbot, 92.6% of participants were confident in using technology, while 7.4% were not. Out of the participants who were interested, a higher percentage were confident in technology (97.1%), while only 2.9% were not. In other words, and perhaps unsurprisingly, people who reported less confidence with technology use in general were also less interested in trying a chatbot for asthma.

Moreover, two Mann-Whitney U tests showed that participants who reported more confidence in their ability to use technology used WhatsApp and other Messenger apps significantly more frequently in comparison to people who feel less confident with technology use (U = 23652.50, Z = -5.636, p <.001) and participants who were interested in using an asthma chatbot also used both WhatsApp and other Messenger apps significantly more frequently in comparison to participants who were not interested in using a chatbot (U = 175969.50, Z = -3.688, p <.001).

We also conducted a CHI Square test for association to measure the relationship between participants' previous use of a virtual assistant and participants' interest in an asthma chatbot. The relationship is significant $\chi2(1) = 10.11$, $p = .001$, $\varphi = .090$. Overall, the majority reported having used a virtual assistant in the past (89.3%). Out of the participants who were not interested in an asthma chatbot, 86.3% had used a virtual assistant before, while 13.7% had not. Of participants who were interested, a higher percentage had used a virtual assistant previously (91.9%) while only 8.1% had not. As such, people who had used a virtual assistant in the past were more likely to be interested in using a chatbot for their asthma.

### 4.9 Lower Confidence in Disease Self-Management Predicts Interest

A CHI Square test for association was conducted to assess the relationship between participants' confidence in the self-management of their asthma and their interest in an asthma chatbot. The relationship is significant $\chi2(2) = 26.84$, $p < .001$, $\varphi = .146$. The majority of participants felt either very confident (51.5%) or somewhat confident (45.6%) about managing their asthma, with only 2.9% of participants who did not feel confident. Out of those interested in the asthma chatbot, 44.7% were very confident, 52.3% were somewhat confident and 3% were not confident at all. In comparison, out of participants who were not interested in an asthma chatbot, a higher percentage indicated being very confident in managing their asthma: 59.1%, 38% were somewhat confident and only 2.9% were not confident. Therefore, participants who are less confident about their own ability to manage their asthma were more likely to be interested in an asthma chatbot. This provides support for the utility of a chatbot in improving management skills and confidence in those who feel they lack them.

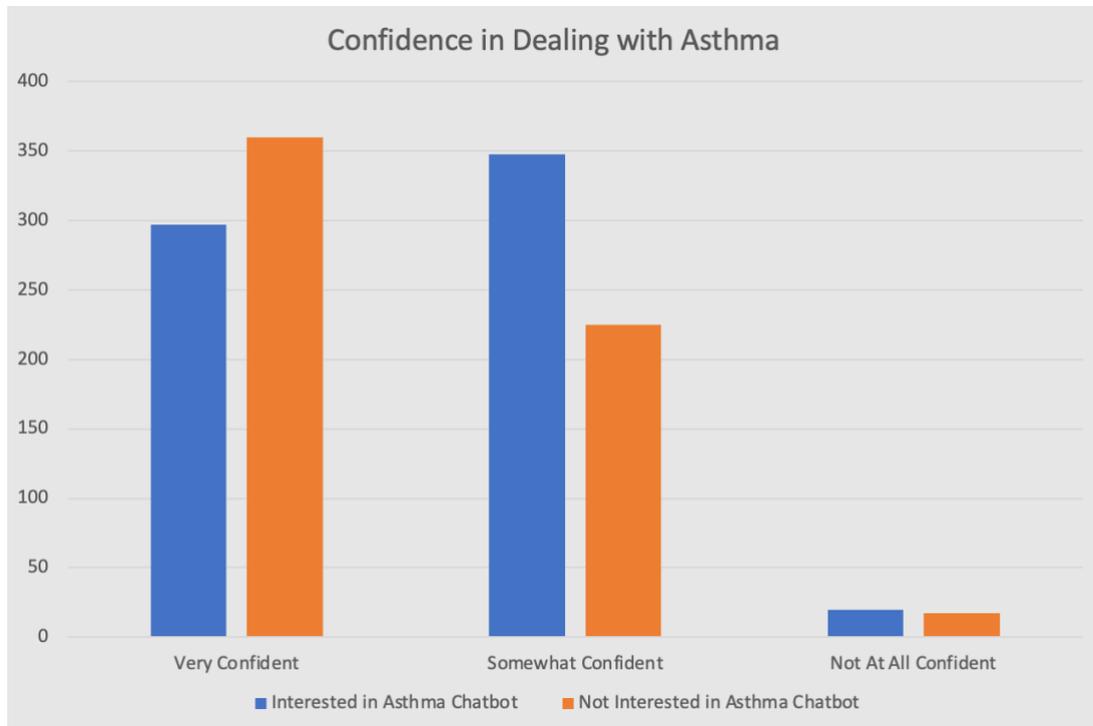

Figure 3: Bar Chart Showing the Confidence Level of Participants Dealing with their Asthma, who are either Interested or not Interested in the Asthma Chatbot.

To assess the relationship between participants' own perception of the severity of their asthma and their interest in using an asthma chatbot, a CHI Square test for association was conducted. The relationship is significant $\chi^2(2) = 16.68$, $p < .001$, $\varphi = .115$. Asthma seriousness was rated by participants, wherein 6.2% rated their asthma as very serious, 31.1% as somewhat serious and 62.7% as not serious. Out of participants who were interested in the asthma chatbot, 6.9% rated their asthma as very serious, 35.6% as somewhat serious and 57.4% as not serious. Participants who were not interested, rated their asthma as less serious overall, with 5.4% indicating that their asthma is very serious, 26% stating that their asthma is somewhat serious and 68.6% rating their asthma was not serious. Overall, participants who were not interested in an asthma chatbot rated their asthma as less serious.

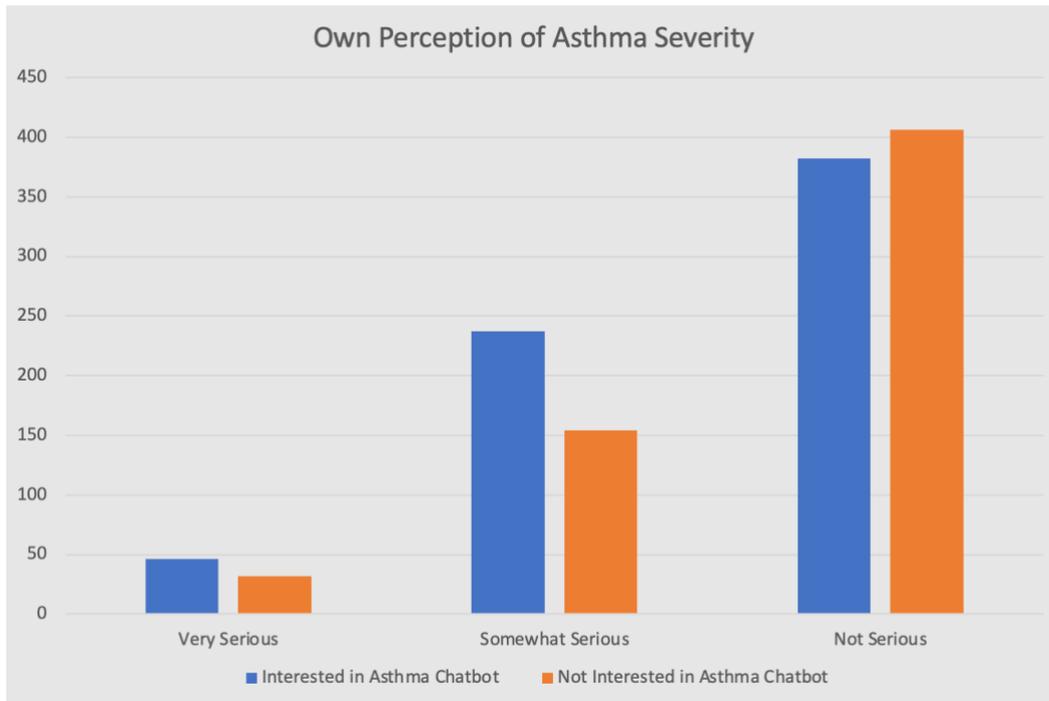

Figure 4: Bar Chart Showing the Level of Severity at which Participants rate their Asthma, who are either Interested or not Interested in the Asthma Chatbot.

Table 4: Summary of Key Results from Each Section.

| | Results |
|---|---|
| **Trust in Healthcare System** | <ul><li>The majority of participants trust the healthcare system (83%).</li><li>People who are interested in the asthma chatbot have more trust in the healthcare system than those who are not interested.</li><li>79.5% of participants seek support from a GP.</li><li>Participants with an interest in the asthma chatbot seek more support from a GP than those who were not interested.</li><li>The more education participants have, the more likely they are to trust the healthcare system.</li><li>Participants who are more confident in their ability to use technology have a higher trust in the healthcare system.</li><li>Older people trust the healthcare system more than younger participants.</li><li>Participants who are already confident in dealing with their own asthma trust the healthcare system more.</li></ul> |
| **Communication Preferences** | <ul><li>59.7% of participants were interested in the voice feature.</li><li>Participants with a general interest in the asthma chatbot were also more interested in the voice feature of the asthma chatbot.</li><li>85.5% of participants would prefer to talk to a human about their asthma.</li><li>Participants who were not interested in the asthma chatbot have a stronger preference for in-person conversations about their asthma.</li></ul> |
| **Conversational Style** | - Most respondents preferred a 'reassuring', 'friendly' and an asthma chatbot that felt 'like talking to a nurse'.<br>- However, a significant number of participants selected other preferences, for example a 'direct' approach or an experience that was more 'like talking to a doctor'.<br>- Overall, it can be concluded that different styles suit different people. |
| **Technology Experience** | - The large majority of participants were confident in their use of technology (95%).<br>- People who are less confident with technology use were also less interested in trying a chatbot for asthma.<br>- Participants who were confident in their use of technology as well as participants who were interested in using an asthma chatbot used WhatsApp and other Messenger Apps significantly more often in comparison to participants who were less confident in their use of technology and those who were not interested in an asthma chatbot.<br>- The majority of participants had used a virtual assistant in the past (89.3%).<br>- People who had used a virtual assistant in the past were more likely to be interested in using a chatbot for their asthma. |
| **Confidence in Self-Management** | - Most participants felt either very confident (51.5%) or somewhat confident (45.6%) about managing their asthma, with only 2.9% of participants who did feel confident.<br>- Participants who were not interested in the asthma chatbot reported to be more confident in managing their asthma.<br>- Seriousness of their asthma was rated by participants, with 6.2% as very serious, 31.1% as somewhat serious and 62.7% as not serious.<br>- Participants who were not interested in the asthma chatbot rated their asthma less serious. |

Table 5: Summary of Results that Distinguish Participants Interested v. Not Interested in an Asthma Chatbot.

| **Interested in an Asthma Chatbot** | **Not interested in an Asthma Chatbot** |
|---|---|
| - More trust in the healthcare system<br>- More likely to ask for support from a GP than those<br>- More likely to also be interested in a feature that detect asthma from sound of voice<br>- Tend to rate their asthma as more serious | - Stronger preference for in-person conversations about their asthma<br>- More confident in managing their asthma |

### 4.10 Qualitative results overview

All of the 1257 survey respondents provided at least one free text response to an open question. While quantitative analysis of closed survey questions reveals characteristics correlated with those most interested in an asthma chatbot, open text results provide explicit insights into the reasons some are interested while others are not. In alignment with this 'interest v. no-interest' focus applied to the quantitative analysis, the qualitative analysis of open question text was also primarily organised around these two categories, including reasons given for these and associated characteristics.

Open text also provided additional information on specific technologies and information sources for answers to survey questions about where participants seek asthma information and what technologies they already use to support self-management.

### 4.11 Reasons for lack of interest and reservations

Data was organised into a set of 4 recurring themes based on the types of reasons and reservations expressed in relation to lower interest in an asthma chatbot. These are presented in Table 6 along with example quotations of respondent comments.

The first theme '**Technology scepticism**' captures expressions of doubt in the accuracy and trustworthiness of chatbot technology. For example, a number of participants expressed scepticism about the capacity of current technology to provide information that is: accurate/safe (referring to the possibility for misdiagnoses and mistakes). Some participants expressed general dislike for technological solutions "I dislike AI" (P9) or referenced the technology's non-humanness as an inherent problem "They're not real…and are not to be taken seriously" (P352) or relatedly, the simple preference for human solutions: "I would rather talk to my doctor or asthma nurse" (P109).

The second recurring theme '**Personalisation scepticism**' refers to responses expressing doubt in a technology's ability to respond in a personally relevant and specific way. For example, Participant 163 states: "I worry that it would tell me basic things that I already know or give generic advice. If it was very tailored to me that might help".

The third recurring reason for lack of interest in a chatbot was **perceived low asthma severity** resulting in a perceived lack of need for technological support, e.g. "I don't feel that my asthma is severe enough to need anything more than my annual check at the moment" (P367)

Finally, a fourth reason for hesitancy in using an asthma chatbot was concern about **system security and data privacy**, including misuse by a malicious third party, e.g. *"The system gets hacked, and my data is used nefariously." (P35)*. For more examples, see Table 6.

Table 6: Summary of reasons for not being interested in an asthma chatbot.

| Theme | Example Quotations |
|---|---|
| **Theme 1: Technology scepticism** | *"I'm unsure it would get details correct and misdiagnose" (P21)*<br><br>*"They're not real, just programs. So they make mistakes and are not to be taken seriously" (P351)*<br><br>*"I dislike AI" (P9)*<br><br>*"I would rather talk to my doctor or asthma nurse" (P109)* |
| **Theme 2: Personalisation scepticism** | *"I'm not sure I would use it; these things tend to give general advice & if you need advice specifically for you then I'm not sure it could do that for everyone" (P231)*<br><br>*"I worry that it would tell me basic things that I already know or give generic advice. If it was very tailored to me that might help" (P163)*<br><br>*"My concern is these virtual assistants are programmed to answer specific questions whilst I may want to ask something that they are not equipped to answer!"* (P455) |
| **Theme 3: Low asthma severity** | *"I don't feel that my asthma is severe enough to need anything more than my annual check at the moment" (P367)*<br><br>*"If my asthma was worse than it is, I would find it more useful." (P855)* |
| **Theme 4: Security and privacy concerns** | *"The system gets hacked, and my data is used nefariously." (P35)*<br><br>*"I'd be uncomfortable with the idea of my medical details being stolen either by a hack or "man in the middle" style attack."(P633)* |

In addition to expressions of doubt, concern, or reservation with respect to using an asthma chatbot, many participants expressed optimism and interest in the potential for a chatbot to be provide benefits. The recurring themes around these responses are summarised in Table 7. They include the capacity for a chatbot to provide **education and self-management support**, for example, by providing users with health information, warning of future asthma attacks, help managing symptoms, and in doing these things, providing an alternative to traditional health services, e.g. "Support for my symptoms without risking a trip to the doctors" (P125); "I have mild asthma and am concerned it may get worse, an Asthma Bot may help me control it because the NHS website is next to useless." (P480).

Another set of responses captured various ways a chatbot provides **ease of access** to support, as compared to traditional health services ("I need help now but cannot see my GP for seven weeks", P428) and with an emphasis by some participants on the importance of such a service being available for free ("If it was free", P107; It being free would be best", P226).

Finally, some participants highlighted the importance of **trusted endorsement** by a trustworthy organisation, such as their GP or the NHS rather than a commercial company, as being important to their interest in using the technology, e.g. "That It is run by the NHS in conjunction with asthma UK not a commercial computer program." (P356)

Table 7: Summary of reasons for interest in an asthma chatbot.

| Theme | Example Quotations |
|---|---|
| **Theme 1: Health Education and self-management support** | *"Help control my asthma" (P120)*<br><br>*"Support for my symptoms without risking a trip to the doctors" (P125)*<br><br>*"To try and predict potential future asthma attacks in case I'm more at risk" (P240)*<br><br>*"It's sounds easy and convenient and help manage my asthma worries easily" (P259)*<br><br>*"Learn new thing about controlling Asthma" (P337)*<br><br>*"It seems like a fun way to learn about the condition." (P453)*<br><br>*"I have mild asthma and am concerned it may get worse, an Asthma Bot may help me control it because the NHS website is next to useless." (P480)* |
| **Theme 2: Ease of access** | *"It already knows the range of symptoms and can come up with something concrete very quickly, instead of me wasting someone's actual time." (P371)*<br><br>*"I need help now but cannot see my GP for seven weeks" (P428)*<br><br>*"it is easier than getting appointment with nurse" (P449)*<br><br>*"I would use it rather than visit A&E out of hours." (P485)*<br><br>*"An easy to access service" (P53)*<br><br>*"If it was free" (P107)*<br><br>*"Free to use and could be helpful before going to see a doctor" (P152)*<br><br>*"It seems like a good tool. It being free would be best" (P226)* |
| **Theme 3: Trusted endorsement** | *"That It is run by the NHS in conjunction with asthma UK not a commercial computer program." (P356)*<br><br>*"Endorsement by my GP practice" (P370)* |

### 4.12 Sources of asthma support: sources and apps

Finally, participants used open text responses to provide the names of information sources and asthma technologies used. Specifically, 74 participants listed specific technologies they already use, or have used in the past, to get support for their asthma. Those best represented in the text responses included the NHS app, (listed by 17 participants), Asthma UK (listed by 9), and Peak Flow (listed by 6).

## 5 DISCUSSION

The current study is the first to assess and contribute detailed insights on the motivational factors, features and characteristics necessary for ensuring an asthma chatbot is of value to adults with asthma and that would maximise the likelihood of meaningful engagement. These findings are based on the closed and free-text responses of adults with asthma sharing views about their expectations, needs and preferences which were then correlated with differences in background, technology confidence, asthma history, and asthma self-management confidence. The goal is to contribute toward meaningful engagement with an easy-access asthma support technology to help address the care gap. Key findings that can most readily guide design and development of future asthma chatbots are summarised herein.

### 5.1 Recommendation 1: Prioritise ease of access and consider existing technologies rather than a custom app

Overall, our results suggest that a significant percentage of UK adults with asthma (53% of our survey participants) would be interested in using a chatbot to support their asthma, and that an even higher number would be interested in using a voice-based risk detection feature (59% of our survey participants). Moreover, our findings provide a strong rationale for delivering the chatbot via WhatsApp instead of through a custom asthma app, since the large majority of our participants had already installed WhatsApp on their phone (85% of 1011 participants) and reported it to be their most common method of sending messages. Finally, WhatsApp was the number one platform of preference for interacting with an asthma chatbot. Alternatively, a website was rated the second most popular channel for accessing an asthma chatbot. Although app have distinct advantages that will sometimes warrant the extra friction, few participants had used an app for asthma and their two top preferences for access leverage familiar and easily accessed technology.

This aligns with qualitative findings in which participants highlighted ease-of-access as a major motivator for using an asthma chatbot. Ease-of-access had three main facets including: a) that it is available 24/7 including when appointments are not; b) that it can (and should) be made available for free to ensure access; and c) that it is easier to access than healthcare professionals.

### 5.2 Recommendation 2: Centre the needs of people with more serious asthma who feel they need additional support.

Our findings indicate that an asthma chatbot is of particular interest to those who believe their asthma to be more serious. As such, a CA could be most beneficial for those patients who need additional support, either because their asthma is indeed more serious, or because they experience more anxiety around it and less confidence around self-management. This is supported by the further results showing that participants who feel less confident in managing their asthma, are also more likely to be interested in a chatbot than those who are already confident. Qualitative data confirms this and suggests that at least some people may also feel they do not always get adequate or timely access to traditional care (see Table 7). For example, participant open responses included: "could be helpful before going to see a doctor" and "I need help now but cannot see my GP for seven weeks". Thus, collectively, our results suggest there is a strong opportunity to integrate conversational agents as a supplementary, easily accessible, 24/7 service for patients who are in need of additional support.

Relatedly, participants who would be interested in using an asthma chatbot, were also more likely to seek support from a GP in comparison to those who were not. This is consistent with the finding that participants interested in a chatbot have more need for support as they rate their asthma is more serious.

In summary, our results show an **asthma chatbot is likely to be of greatest benefit to**:
1. Patients who believe their asthma to be serious.
2. Patients who feel less confident about managing their asthma.
3. Patients looking for immediate support while they wait for an appointment with a healthcare professional.

### 5.3 Recommendation 3: Include conversational content for asthma education, risk assessment and customised self-management advice.

Across both quantitative and qualitative responses, results demonstrate that participants would value easy access to educational information about asthma, to support in assessing and preventing asthma attacks and support for self-management, in particular, tailored support that goes beyond the generic advice available broadly. The clinicians involved in this study add that there is great the potential for a chatbot to have conversations about topics that are important to positive health outcomes but that there often isn't enough time to cover during medical appointments. These topics include full explanations of what asthma is, what the medicine is for, how and why it works, proper technique, and conversations to support identifying and avoiding triggers. This would not only fill in gaps, but provide repeated access to the information over time, thus augmenting clinical practice.

### 5.4 Recommendation 4: Craft a conversational design that is reassuring and like talking to a nurse

Our survey results with respect to people's preferences for the conversational style (personality) of a chatbot revealed a set of majority preferences with additional groups of alternate preferences. Most respondents preferred an asthma chatbot that was 'reassuring', friendly' and 'like talking to a nurse. It's notable however, that a significant alternative group favoured a 'direct' approach or an experience that was more 'like talking to a doctor'. Designers and developers could approach this diversity in a number of ways, including by a) opting for the majority preference, b) targeting a specific audience or c) allowing users to select personalities from a set of options.

### 5.5 Recommendation 5: Address security and privacy concerns

The group of participants that reported more interest in an asthma chatbot also showed more overall trust in the healthcare system in comparison to participants who were not interested. This finding would suggest that participants who are not interested in the chatbot might be overall less inclined to seek external support for their asthma. Future studies could investigate whether this is simply due to them having less serious asthma and effective self-management, or a general lack of trust in external support (e.g., negative experience, language barriers etc.). The qualitative analysis provided some evidence for both explanations indicating a combination of reasons is likely. Specifically, on the one hand, we identified a group of participants that were not interested in the chatbot and believed their asthma to be less severe (see Table 6, Theme 3). On the other hand, we also identified a group that were sceptical of AI solutions and their accuracy in general.

Qualitative results revealed a number of themes that provide richer explanations for the quantitative findings. Specifically, that there were at least four categories of reasons given for lack of interest in an asthma chatbot:

1. **Technology scepticism** (doubt in the technologies ability to be accurate or useful)
2. **Personalisation scepticism** (doubt in the technology's ability to provide anything other than very basic and generic advice or to provide specific advice tailored to specific needs)
3. **Low asthma severity** (perception that their asthma isn't severe enough to warrant the use of technological support)
4. **Security and privacy concerns** (concerns over data misuse or hacking)

A successful chatbot would have to address these concerns to reach ambivalent users. These patient concerns demonstrate the importance of providing information and transparency around data security and the technology owner. Moreover, as the results show that previous exposure to a virtual assistant had a positive impact on patients' interest in the CA, these issues may change over time as exposure increases.

### 5.6 Recommendation 6: Address technology scepticism.

While security and privacy concerns have to do with threats to personal autonomy, technology scepticism was evidenced by participant responses to do with non-belief in the technology's capabilities, accuracy, and fitness to purpose. Some participants expressed aversion to using technology in favour of human interaction or general aversion to "AI". Others were doubtful that a chatbot system could provide advice that was tailored enough to the individual to be useful. Aiming to address these doubts, both technologically (making the doubts unfounded) and/or through messaging (communicating the genuine benefits and limitations) is likely to increase the audience for an asthma chatbot.

However, trust in these systems is likely to be a moving target. Given the rapid growth of generative AI chatbots like ChatGTP, future studies could explore whether increased exposure to these types of chatbots impacts either positively or negatively on trust in related technologies. Although an asthma chatbot should only provide advice vetted by clinicians, generic generative AI chatbots like ChatGPT and Bard generate advice that is not safety checked by a human and may therefore respond erroneously with so-called "hallucinations" or fabrications, which, in a health context, could be highly dangerous. As such, educating patients as to the limitations and distinctions among various conversational technologies may become critical going forward.

## 5.7 Recommendation 7: Consider a trusted endorsement.

A final point on trust came from participants who highlighted the importance that a chatbot be endorsed by trusted authorities such as their GP, an asthma charity and not a commercial enterprise.

By understanding people's doubts and the reservations that form obstacles to technology adoption, developers can target these concerns through technological improvements, interaction design and education.

## 6 LIMITATIONS AND FUTURE WORK

A limitation to the generalisability of the findings within this study is that it was conducted within the UK, and with the UK healthcare system in mind. Results with respect to access to primary care, trust in the healthcare system, or access to technologies, for example, are likely to vary across other countries. Another important consideration is that the survey was conducted online through an online survey service. Confidence in technology use was reported as very high within our sample, but results could be different, for example, if a paper-based survey was conducted via community centres.

It is also relevant to highlight that our data relies on participant self-report of having asthma (rather than medical proof). To mitigate the risk of inaccurate self-reporting, we asked participants to specify when they were diagnosed by a doctor and to answer questions on symptoms and asthma control. We believe the added detail in questions combined with the high number of participants who took part in the study, mitigates potential effects of inaccurate self-reporting.

Another limitation involves our small sample of participants who identify with a minority ethnic group. We did not recruit enough participants from an ethnic minority group through the YouGov database to be able to explore research questions that could contribute to helping these groups get their needs met. We believe that this is a worthwhile research area for future studies since there are often health disparities demonstrated for minority ethnic communities and easily accessible technologies, such as conversational agents, might contribute to addressing some of these.

Relatedly, the majority of participants in our sample gets support from a GP for their asthma and trust in the healthcare system was high. Follow-up studies could focus specifically on patient groups who do not fall within either of these categories in order to gain a better understanding of how these groups (who are not accessing traditional care) might be better supported and how a chatbot might play a role.

Lastly, in the current study, participants did not have the chance to test an actual asthma chatbot. Instead, they were asked to make hypothetical judgments about their interests and preferences, so their responses were speculative and based on their previous experiences. In a follow-up study we will test the feasibility of a prototype designed in alignment with the findings of this study. The follow-up study will provide insights into real-world patterns of use, perception of benefit, and final evaluations.

## 7 CONCLUSION

Overall, the current study provides evidence for demographically broad interest in an asthma chatbot among adults with asthma in the UK. Moreover, we were able to identify key factors that correlate with greater interest (i.e. easy access, fast alternative to GP/nurse, lower confidence in asthma self-management, perception of asthma as serious) and less interest (lack of trust in AI, security concerns, preference for in-person contact, higher confidence in self-management). These results offer important insights for the design and development of a CA for asthma management (targeting the needs of those most interested and the concerns of those less interested) and demonstrate the demand and opportunity in the UK for providing alternative support for asthma management that can be easily accessed outside of appointments with a healthcare professional and to help improve risk self-assessment, asthma control and access to basic care.

## 8 ACKNOWLEDGEMENTS

This project was funded by the UK's Engineering and Physical Sciences Research Council [grant EP/W002477/1]. We are grateful to Asthma+Lung UK and our patient advisory board participants for their constant support. This research was supported by the NIHR Imperial Biomedical Research Centre (BRC).